\documentclass[journal]{IEEEtran}
\usepackage{cite}
\ifCLASSINFOpdf
  \usepackage[pdftex]{graphicx}
  \DeclareGraphicsExtensions{.pdf,.jpeg,.png}
\else
\fi
\usepackage{subcaption}

\usepackage{amsmath}
\usepackage{amssymb}
\usepackage{gensymb}

\usepackage{array}
\usepackage{booktabs}

\begin{document}

\title{Performance Analysis of AFDM Under In-Phase and Quadrature Imbalance at Receiver}
% \title{BER Analysis of AFDM Systems With Receiver IQ Imbalance }
% 


\author{Zhenfeng~Huang,
Yitong Liu,
Hongwen Yang
        % Michael~Shell,~\IEEEmembership{Member,~IEEE,}
        % John~Doe,~\IEEEmembership{Fellow,~OSA,}
        % and~Jane~Doe,~\IEEEmembership{Life~Fellow,~IEEE}
% \thanks{M. Shell was with the Department
% of Electrical and Computer Engineering, Georgia Institute of Technology, Atlanta,
% GA, 30332 USA e-mail: (see http://www.michaelshell.org/contact.html).}% <-this % stops a space
% \thanks{J. Doe and J. Doe are with Anonymous University.}% <-this % stops a space
% \thanks{Manuscript received April 19, 2005; revised August 26, 2015.}
}

\maketitle

\begin{abstract}
Affine Frequency Division Multiplexing (AFDM) is a chirp-based multicarrier waveform that achieves full diversity in doubly selective channels while requiring reduced pilot overhead. 
It is regarded as a highly promising candidate for sixth-generation (6G) mobile communication waveforms in high-mobility scenarios. 
However, AFDM deployment remains subject to hardware impairments, particularly the in-phase and quadrature (IQ) imbalance commonly encountered in direct conversion transceivers.
% At present, the impact of IQ imbalance on OFDM and OTFS systems has been investigated in the literature; however, its effect on the performance of AFDM systems has not yet been sufficiently explored.
This paper investigates the impact of receiver IQ imbalance on the bit error rate (BER) performance of AFDM systems. 
A mathematical model of AFDM under receiver IQ imbalance is first established, where the resulting inter-carrier interference (ICI) in the discrete affine Fourier transform (DAFT) domain is explicitly characterized. 
Moreover, a closed-form expression for the BER is derived under the influence of receiver IQ imbalance in an M-QAM-AFDM system over an AWGN channel.
Numerical simulation results validate the accuracy of the theoretical analysis, while also indicating that under identical IQ imbalance conditions, AFDM exhibits more pronounced BER degradation compared to OFDM. 
% , with this effect becoming more pronounced at higher modulation orders.
% This study provides a theoretical basis for evaluating the performance of AFDM systems under practical hardware impairments.
The results provide fundamental insights into the sensitivity of AFDM to receiver IQ imbalance and offer guidance for practical system design. 
\end{abstract}


% \fi
% \IEEEpeerreviewmaketitle

\section{Introduction}
% for IEEE journal papers produced under \LaTeX\ using
% IEEEtran.cls version 1.8b and later.
% % You must have at least 2 lines in the paragraph with the drop letter
% % (should never be an issue)
% I wish you the best of success.

 

\IEEEPARstart{C}{urrently}, space-air-ground integrated wireless networks have become the main trend for next-generation mobile communication networks, with low-Earth orbit satellite communications poised to become a crucial component of 6G \cite{xiao_space-air-ground_2024}, \cite{zhang_space-air-ground_2025}, \cite{cui_space-air-ground_2022}. 
In such highly dynamic environments, conventional multicarrier waveforms suffer from performance degradation due to Doppler dispersion, motivating the investigation of alternative waveform designs. 
% However, the traditional waveforms currently employed in mobile communication systems are primarily based on OFDM. In highly dynamic scenarios such as LEO satellite communications within 6G, OFDM will be susceptible to Doppler frequency shifts, leading to severe inter-carrier interference (ICI) and ultimately resulting in a significant degradation of transmission performance \cite{liu_otfs_2025}.
% To address this issue, Hadani et al. proposed orthogonal time frequency space (OTFS) modulation in \cite{hadani_orthogonal_2017}. 
% This approach places symbols in the delay-Doppler (DD) domain, converting the fading, time-varying wireless channel experienced by the modulated signal into a time-independent channel with a complex channel gain that remains roughly constant across all symbols. This enables full utilization of the complete diversity in both time and frequency \cite{yi_hong_delay-doppler_2022}. 
% However, OTFS incurs considerable pilot overhead due to its two-dimensional structure.

Recent studies indicate that AFDM exhibits a superior diversity order in doubly dispersive channels and outperforms orthogonal time frequency space (OTFS) modulation in terms of pilot overhead, making it a strong candidate waveform for high-mobility scenarios in future wireless systems \cite{benzine_affine_2023}, \cite{yin_affine_2025}, \cite{bemani_affine_2023}, \cite{wang_afdm_2025}.
AFDM is a novel chirp-based multi-carrier waveform, which multiplexes symbols in the discrete affine Fourier transform (DAFT) domain. Taking advantage of the chirp operation inherent in DAFT, by appropriately adjusting parameters $c_1$ and $c_2$, paths with different delays can be distinguished in the DAFT domain without overlap, thus achieving full diversity in doubly dispersive channels \cite{bemani_afdm_2021}.

However, challenges in the practical implementation of AFDM in real-world scenarios remain to be analyzed, particularly when confronted with the impact of hardware impairments. 
% Due to its transformation generalization from OFDM, AFDM suffers from a more severe peak-to-average power ratio (PAPR) problem than OTFS \cite{rou_orthogonal_2024}, which severely limits its output power under finite linear-region power constraints. Several studies have been conducted to address this issue.
% Tao \textit{et al.} proposed a DAFT-Spread AFDM scheme in \cite{tao_daft-spread_2024} and \cite{tao_affine_DAFT-S_2025}, which incorporates an additional parameter-flexible DAFT operation prior to AFDM modulation. By appropriately selecting the transform parameters, this approach can significantly reduce PAPR.
% Reddy and Bitra introduced the normalized $\mu$-law companding transformation in \cite{reddy_papr_2025}, which can reduce PAPR by compressing the peak values of AFDM signals while maintaining constant average power. However, this approach incurs a loss in BER performance.
% Yuan \textit{et al.} proposed a grouped pre-chirp selection algorithm in \cite{yuan_papr_2025}. At the cost of lower communication performance, AFDM adopting the proposed algorithm can achieve better PAPR performance than OTFS while maintaining lower modulation complexity.
In direct conversion transceivers that leverage low cost as a key advantage, manufacturing variations typically introduce gain and phase mismatches between the in-phase (I) and quadrature (Q) channels that transmit the oscillator signal to the mixer, leading to IQ imbalance or IQ mismatch (IQmm) \cite{neelam_channel_2021}.
IQ imbalance is a common hardware impairment that disrupts the orthogonality between the I and Q channels, introducing mirror interference and signal distortion. This degrades the system's BER performance, with the adverse effects being further amplified in high-dynamic, low-SNR scenarios \cite{sandell_estimation_2021}.

Regarding the impact of IQ imbalance on OFDM systems, Zareian and Vakili presented a BER performance analysis of M-QAM-OFDM systems under IQ imbalance conditions in \cite{zareian_analytical_2007}. The authors in \cite{aziz_comparative_2022}, \cite{nishibe_novel_2022}, and \cite{sun_low-complexity_2019} have all proposed compensation algorithms for IQ imbalance in OFDM systems.
As for OTFS, Tusha \textit{et al.} analyzed the impact of IQ imbalance in the delay-Doppler domain in \cite{tusha_physical_2021}, revealing the mirror Doppler interference and power degradation effects caused by IQ imbalance. In \cite{tusha_performance_2021}, they examined the BER performance of OTFS under transmitter IQ imbalance and derived a closed-form expression for BER under ideal channel estimation conditions.
Bhagat \textit{et al.} investigated the impact of receiver IQ imbalance on BER performance in OTFS systems assisted by an intelligent reflecting surface in \cite{bhagat_ber_2024}.
Neelam and Sahu proposed an estimation and compensation algorithm for direct current (DC) offset and IQ imbalance in OTFS systems in \cite{neelam_analysis_2022}, which first performs DC offset and channel estimation followed by IQ imbalance estimation and compensation. The proposed method effectively mitigates the impact of DC offset and IQ imbalance on OTFS systems.

Recently, several studies have begun to investigate the impact of hardware impairments on AFDM systems. For example, Gunasekara and Bedeer \cite{gunasekara_analysis_2025} analyzed the effects of receiver IQ imbalance and residual carrier frequency offset in AFDM systems under doubly selective channels. Their study shows that the complex-conjugate operation inherent in the IQ imbalance model may destroy the sparsity of the effective AFDM channel matrix in the presence of receiver IQ imbalance. To mitigate the resulting performance degradation, an improved LMMSE detector capable of handling improper Gaussian noise was proposed.
In addition, Sui \textit{et al.} \cite{sui_mimo-afdm_2026} investigated the joint impact of multiple hardware impairments on MIMO-AFDM systems and developed analytical performance results under various hardware non-idealities. However, their work mainly focuses on transmitter IQ imbalance and the combined effect of multiple impairments, while the statistical characteristics of the interference introduced by receiver IQ imbalance and its impact on the error-rate performance have not been explicitly characterized.

In this paper, we focuse on the statistical characterization of the interference caused by receiver IQ imbalance under the AFDM modulation structure and its impact on the error-rate performance. 
Specifically, a mathematical model of AFDM systems with receiver IQ imbalance is established under the AWGN channel, and the statistical distribution of the resulting interference is analyzed. Based on these results, a closed-form expression of theoretical BER is obtained for M-QAM AFDM systems in the presence of receiver IQ imbalance. 
A comparative study of BER performance between AFDM and OFDM systems is conducted under identical IQ imbalance conditions.  The results indicate that AFDM exhibits a greater sensitivity to IQ imbalance at the receiver end compared to OFDM, with this effect becoming particularly severe under higher-order modulation schemes.
% The analysis provides new insights into the mechanisms through which IQ imbalance affects the error-rate performance of AFDM systems.

% The main contributions of this paper are as follows.
% \begin{itemize}
%     \item The IQ imbalance issue at the receiver end of AFDM systems is modeled, its impact in the DAFT domain is analyzed, and the inter-carrier interference (ICI) induced by receiver IQ imbalance is revealed.
%     \item By approximating the ICI from multiple carriers as a Gaussian distribution, the BER performance of an AFDM system affected by IQ imbalance at the receiver under an AWGN channel is derived, yielding a closed-form expression.
%     \item A comparative test of BER performance between AFDM and OFDM systems is conducted under identical IQ imbalance conditions.  The results indicate that AFDM exhibits a greater sensitivity to IQ imbalance at the receiver end compared to OFDM, with this effect becoming particularly severe under higher-order modulation schemes.
% \end{itemize}

The rest of the paper is organized as follows. Section II presents the system model of AFDM under receiver IQ imbalance. Section III analyzes the impact of receiver IQ imbalance on AFDM systems and characterizes the statistical properties of the resulting noise and interference terms. In Section IV, a closed-form BER expression for AFDM systems under receiver IQ imbalance is derived. Numerical simulation results are provided in Section V to verify the theoretical analysis. Finally, Section VI concludes the paper. 

\textit{Notations}: 
Upper-case boldface letter $\textbf{A}$ and lower-case boldface letter $\textbf{a}$ denote a matrix and a vector, respectively.
$\textbf{A}_{m,n}$ denotes the element in the $m$-th row and the $n$-th column of $\textbf{A}$.
$\textbf{I}_N$ denotes the $N\times N$ identity matrix.
$\left( \cdot \right) ^T$,  $\left( \cdot \right) ^*$, and $\left( \cdot \right) ^H$denote transpose, complex conjugate, and Hermite transpose operations, respectively. 
$x \sim \mathcal{N} \left( \mu _{x},\sigma _{x}^{2}\right) $ denotes that the real random variable $x$ follows a normal distribution with mean $\mu _{x}$ and variance $\sigma _{x}^{2}$.
$z \sim \mathcal{C} \mathcal{N} \left( \mu _{z},\sigma _{z}^{2},\kappa _{z} \right) $ denotes that the complex random variable $z$ follows a complex Gaussian distribution with mean $\mu _{z}$, variance $\sigma _{z}^{2}$ and pseudo-variance $\kappa _{z}$.
$j$ denotes the imaginary unit.

\section{System Model}
This section presents the mathematical model of an AFDM system under IQ imbalance at the receiver, providing an essential background for subsequent discussions.
\subsection{IQ Imbalance Model}
\begin{figure}[!t]
    \centering
    \includegraphics[width=1\linewidth]{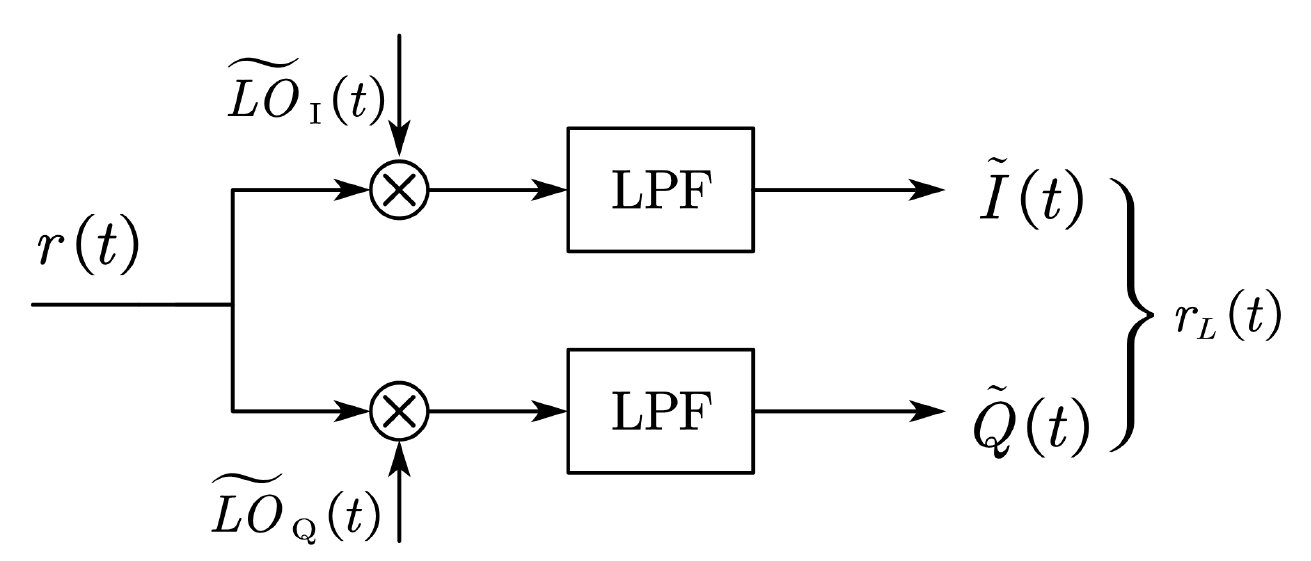}
    \caption{The system model of receiver IQ imbalance.}
    \label{fig:IQmm_model}
\end{figure}

The system model of receiver IQ imbalance is shown in Fig. \ref{fig:IQmm_model}. 

Under ideal conditions, the I and Q carrier signals in a direct conversion receiver are amplitude-matched and phase-orthogonal. However, when IQ imbalance occurs, the expressions for the two carrier signals are as follows \cite{Razavi2012RF}
\begin{equation}
    \left\{ \begin{aligned}
	\widetilde{LO}_I\left( t \right) &=2\left( 1+\varepsilon \right) \cos \left( \omega _ct+\varphi \right)\\
	\widetilde{LO}_Q\left( t \right) &=-2\left( 1-\varepsilon \right) \sin \left( \omega _ct-\varphi \right)\\
\end{aligned} \right.
\end{equation}
where $\epsilon$ is the amplitude imbalance parameter and $\varphi$ is the phase imbalance parameter.
Under IQ imbalance, the amplitude of the two signals deviates and their phases are no longer perfectly orthogonal.
Let the received signal $r(t)$ be demodulated by an ideal receiver to generate the baseband signal $r_L\left( t \right) $, and by a receiver affected by IQ imbalance to generate the baseband signal $\tilde{r}_L\left( t \right) $. 
An IQ imbalance model can be derived in the following form \cite{Razavi2012RF}
\begin{equation}
    \tilde{r}_L\left( t \right) =\alpha r_L\left( t \right) +\beta r_{L}^{*}\left( t \right) ,
\end{equation}
where the two parameters $\alpha$ and $\beta$ are
\begin{equation}
    \left\{ \begin{array}{c}
	\alpha =\cos \varphi +j\varepsilon \sin \varphi\\
	\beta =\varepsilon \cos \varphi -j\sin \varphi\\
\end{array} \right. 
\end{equation}
The ideal case corresponds to $\alpha=1$ and $\beta=0$.

\subsection{AFDM Fundamentals}

% \subsubsection{AFDM Modulation}
Denote $\mathbf{x}\in\mathcal{C} ^{N\times 1}$ by the data symbol in the DAFT domain, and $N$ denotes the number of subcarriers, then the modulated symbol of $\mathbf{x}$ after IDAFT is
\begin{equation}
    s\left[ n \right] =\sum_{m=0}^{N-1}{x\left[ m \right] \phi \left[ n,m \right]}, n=0,\cdots ,N-1,
    \label{eq:AFDMmod_elem}
\end{equation}
where $\phi \left[ n,m \right] $ denotes the time-domain sequence of the $m$th carrier,
\begin{equation}
    \phi \left[ n,m \right] =\frac{1}{\sqrt{N}}e^{j2\pi \left( c_1n^2+c_2m^2+\frac{nm}{N} \right)}
\end{equation}
where $m,n=0,\cdots N-1$. The two adjustable chirp signal parameters $c_1$ and $c_2$ are referred to as the post-chirp parameter and the pre-chirp parameter, respectively \cite{liu_pre-chirp-domain_2025}. 

Note that equation (\ref{eq:AFDMmod_elem}) can be expressed in matrix form as
\begin{equation}
    \mathbf{s}=\mathbf{A}^H\mathbf{x}=\mathbf{\Lambda }_{c_1}^{H}\mathbf{F}^H\mathbf{\Lambda }_{c_2}^{H}\mathbf{x},
\end{equation}
where the diagonal matrix $\mathbf{\Lambda }_c=\mathrm{diag}( e^{-j2\pi cn^2}, n=0,\cdots ,N-1) $; $\mathbf{F}$ denotes the $N$-point DFT matrix with $\mathbf{F}_{m,n}=e^{-j2\pi mn/N}/\sqrt{N}$; $\mathbf{A}=\mathbf{\Lambda }_{c_2}\mathbf{F\Lambda }_{c_1}$ denotes the $N$-point DAFT matrix with $\mathbf{A}_{n,m}=\phi ^*\left[ m,n \right]=e^{-j2\pi \left( c_2n^2+c_1m^2+mn/N \right)}/\sqrt{N}$, and $\mathbf{A}^H\mathbf{A}=\mathbf{AA}^H=\mathbf{I}_N$.

Different from OFDM, AFDM employs a chirp-periodic prefix (CPP) instead of a cyclic prefix (CP) to overcome multipath propagation and mitigate inter-symbol interference (ISI) \cite{bemani_affine_2023}.
A CPP of length $L_{\mathrm{cpp}}$ is defined as
\begin{equation}
    s\left[ n \right] =s\left[ n+N \right] e^{-j2\pi c_1\left( N^2+2Nn \right)}, 
\end{equation}
where $n=-L_{\mathrm{cpp}},\cdots ,-1$. 

\begin{figure*}[!t]
    \centering
    \includegraphics[width=0.75\linewidth]{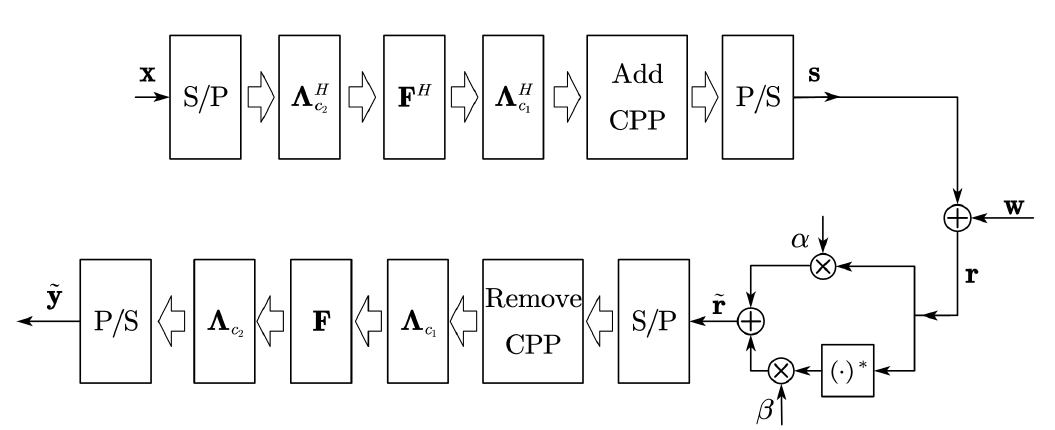}
    \caption{Block diagram of AFDM system affected by receiver IQ imbalance.}
    \label{fig:AFDM_rx_IQmm}
\end{figure*}

In order to achieve the optimal diversity order under a dual-selective channel, the AFDM parameter $c_2$ should be set to an irrational number or a rational number sufficiently smaller than $\frac{1}{2N}$, while the parameter $c_1$ should be set to
\begin{equation}
    c_1=\frac{2\left( \alpha _{\max}+\xi _{\nu} \right) +1}{2N},
    \label{eq:AFDM_c_1}
\end{equation}
where $\alpha _{\max}$ denotes  the maximum integer Doppler frequency shift in the scenario, $\xi _{\nu}$ denotes  an integer that measures the degree of energy dispersion caused by fractional Doppler frequency shifts, as shown in \cite{bemani_affine_2023}.

In the following sections, the parameter $c_2$ is set to an irrational number and the parameter $c_1$ is determined according to equation (\ref{eq:AFDM_c_1}), which means that $2Nc_1$ is a positive odd integer.
% The number of subcarriers $N$ is an integer power of 2.

\subsection{AFDM Under Receiver IQ Imbalance}
This subsection establishes the input/output relationship of an AFDM system under IQ imbalance at the receiver end. The system model is shown in Fig. \ref{fig:AFDM_rx_IQmm}.
The received signal over an AWGN channel is given by
\begin{equation}
    \mathbf{r}=\mathbf{s}+\mathbf{w},
\end{equation}
% where $\textbf{s}$ denotes the signal vector in the time domain, $\textbf{w}$ denotes the noise vector in the time domain and $\mathbf{w}\sim \mathcal{C} \mathcal{N} \left( \mathbf{0},\sigma _{n}^{2}\mathbf{I}_N, \mathbf{0} \right) $.
where $\mathbf{s}\in\mathbb{C}^{N}$ denotes the transmitted signal vector, and $\mathbf{w}\sim \mathcal{C} \mathcal{N} \left( \mathbf{0},\sigma _{n}^{2}\mathbf{I}_N, \mathbf{0} \right) $ is a circularly symmetric complex Gaussian noise vector whose elements are i.i.d.
Due to IQ imbalance at the receiving end, the received sequence becomes
\begin{equation}
    \tilde{\mathbf{r}}=\alpha \mathbf{r}+\beta \mathbf{r}^*=\alpha \left( \mathbf{A}^H\mathbf{x}+\mathbf{w} \right) +\beta \left( \mathbf{A}^T\mathbf{x}^*+\mathbf{w}^* \right) .
\end{equation}

% \subsubsection{AFDM Demodulation}
After the DAFT transformation through AFDM demodulation, the received sequence obtained is
\begin{equation}
\begin{aligned}
    \tilde{\mathbf{y}}&=\mathbf{A}\tilde{\mathbf{r}}\\
&=\alpha \mathbf{A}\left( \mathbf{A}^H\mathbf{x}+\mathbf{w} \right) +\beta \mathbf{A}\left( \mathbf{A}^T\mathbf{x}^*+\mathbf{w}^* \right) \\
&=\alpha \mathbf{x}+\beta \mathbf{AA}^T\mathbf{x}^*+\alpha \mathbf{Aw}+\beta \mathbf{Aw}^*\\
&=\alpha \mathbf{x}+\beta \mathbf{\Xi x}^*+\mathbf{v},
\end{aligned}
\label{eq:AFDM_IQmm_y_mtxform}
\end{equation}
where $\mathbf{v}=\alpha \mathbf{Aw}+\beta \mathbf{Aw}^*$ is the noise vector in the DAFT domain. 
$\mathbf{\Xi }\triangleq \mathbf{AA}^T=\mathbf{\Lambda }_{c_2}\mathbf{F\Lambda }_{c_1}\mathbf{\Lambda }_{c_1}\mathbf{F\Lambda }_{c_2}$, whose element values are
\begin{equation}
    \mathbf{\Xi }_{m,n}=\frac{1}{N}e^{-j2\pi c_2\left( m^2+n^2 \right)}\sum_{k=0}^{N-1}{e^{-j2\pi \left[ 2c_1k^2+\frac{k\left( m+n \right)}{N} \right]}},
    \label{eq:XI_mn}
\end{equation}
where we define $S\left( m+n \right) =\sum_{k=0}^{N-1}{e^{-j2\pi \left[ 2c_1k^2+\frac{k\left( m+n \right)}{N} \right]}}$. Given that $d\triangleq 2Nc_1$ is a positive odd number, the sum $S\left( a \right) =\sum_{k=0}^{N-1}{e^{-j\frac{2\pi}{N}\left( dk^2+ak \right)}}$ can be expressed in the following closed-form formula
\begin{equation}
    S\left( a \right) =\begin{cases}
	0,&	a\,\mathrm{is} \,\mathrm{odd}\\
	\left( 1-j \right) \gamma _{d,N}\cdot \sqrt{N}e^{j\frac{2\pi}{N}\left( \frac{a}{2} \right) ^2d_{-1}},&a\,\mathrm{is}\, \mathrm{even}\\
\end{cases}
\label{eq:S_a}
\end{equation}
where $d_{-1}$ is the modular multiplicative inverse of $d$ modulo $N$, which satisfies $d\cdot d_{-1}\equiv 1 \left( \mathrm{mod} \, N \right) $, $x\equiv y\,\,\left( \mathrm{mod} z \right) $ indicates that the result of taking $x$ modulo $z$ is $y$.
The rotation factor $\gamma _{d,N}$ is related to $d$ and $N$, and its values are determined in different cases as follows
\begin{equation}
    \gamma _{d,N}=\begin{cases}
	1&		d\equiv 1 \left( \mathrm{mod} 8 \right)\\
	j\cdot \left( -1 \right) ^{\log _2N}&		d\equiv 3 \left( \mathrm{mod} 8 \right)\\
	\left( -1 \right) ^{\log _2N}&		d\equiv 5 \left( \mathrm{mod} 8 \right)\\
	j&		d\equiv 7 \left( \mathrm{mod} 8 \right) .\\
\end{cases}
\end{equation}
Note that equation (\ref{eq:AFDM_IQmm_y_mtxform}) can be expressed in an element-wise form as follows
\begin{equation}
    \tilde{y}\left[ m \right] =\alpha x\left[ m \right] +\beta \sum_{m^{\prime} =0}^{N-1}{\xi _m\left[ m^{\prime} \right] x^*\left[ m^{\prime} \right]}+v\left[ m \right] , 
    \label{eq:AFDM_IQmm_y_eleform}
\end{equation}
where the subcarrier index $m=0,\cdots ,N-1$, the coefficient $\xi _m\left[ m^{\prime} \right] $ is the element $\mathbf{\Xi }_{m,m^{\prime}}$ in the $m$-th row and $m^{\prime}$th column of the matrix $\mathbf{\Xi }$.

\section{Impact of Receiver IQ Imbalance on AFDM Systems}
In the mathematical model of  (\ref{eq:AFDM_IQmm_y_eleform}), since the IQ imbalance parameters typically vary slowly over time, their estimation is considerably less challenging than channel estimation in AFDM systems. Therefore, it is reasonable to assume that the parameters $\alpha$ and $\beta$ (or equivalently $\epsilon$ and $\varphi$) are known prior to symbol detection at the receiver.
The receiver can correct the amplitude and phase of the signal item prior to making the symbol decision. The decision variable is
\begin{equation}
    \begin{aligned}
\hat{y}\left[ m \right] &=x\left[ m \right] +\frac{\beta}{\alpha}\sum_{m^{\prime}=0}^{N-1}{\xi _m\left[ m^{\prime} \right] x^*\left[ m^{\prime} \right]}+\frac{1}{\alpha}v\left[ m \right] 
\\
&=x\left[ m \right] +{\frac{\beta}{\alpha}\xi _m\left[ m \right] x^*\left[ m \right] }
\\ &\quad+{\frac{\beta}{\alpha}\sum_{m^{\prime}=0, m^{\prime}\ne m}^{N-1}{\xi _m\left[ m^{\prime} \right] x^*\left[ m^{\prime} \right]}}+\hat{v}\left[ m \right] 
    \label{eq:AFDM_IQmm_y_eleform_sep}
   \\
   &=x\left[ m \right] +e\left[ m \right] +I_{\mathrm{ICI}}+\hat{v}\left[ m \right] ,
    \end{aligned}
\end{equation}
where  $x\left[ m \right]$ denotes the data symbol on the $m$-th subcarrier at the transmitter,  $e[m]\triangleq {\frac{\beta}{\alpha}\xi _m\left[ m \right] x^*\left[ m \right] }$ represents the bias introduced by the data symbol on the current subcarrier,  $I_{\mathrm{ICI}}\triangleq{\frac{\beta}{\alpha}\sum_{m^{\prime}=0, m^{\prime}\ne m}^{N-1}{\xi _m\left[ m^{\prime} \right] x^*\left[ m^{\prime} \right]}}$ denotes the interference caused by the data symbols on all other subcarriers and $\hat{v}\left[ m \right] $ is the equivalent noise.

Consider a rectangular QAM constellation with energy normalization. Let $x\left[ m \right] =x_{\mathrm{I}}\left[ m \right] +jx_{\mathrm{Q}}\left[ m \right] $, where the subscripts I and Q denote the real and imaginary parts, respectively. Decisions are made separately for the I and Q channels, with the decision inputs given by
\begin{subequations}
  \begin{align}
    \hat{y}_{\mathrm{I}}\left[ m \right] &=x_{\mathrm{I}}\left[ m \right] +e_{\mathrm{I}}\left[ m \right] +I_{\mathrm{I}}^{\mathrm{ICI}}+\hat{v}_{\mathrm{I}}\left[ m \right] ,\\
    \hat{y}_{\mathrm{Q}}\left[ m \right] &=x_{\mathrm{Q}}\left[ m \right] +e_{\mathrm{Q}}\left[ m \right] +I_{\mathrm{Q}}^{\mathrm{ICI}}+\hat{v}_{\mathrm{Q}}\left[ m \right] .
  \end{align}
\end{subequations}

In the following subsections, we will analyze the distributions of the noise component $\hat{v}\left[ m \right] $ and the interference component $I_{\mathrm{ICI}}$, respectively.

\subsection{Distribution of the Noise Component}
The equivalent noise term in equation (\ref{eq:AFDM_IQmm_y_eleform_sep}) can be expressed as follows
\begin{equation}
    \hat{v}\left[ m \right] =\sum_{m^{\prime}=0}^{N-1}{\mathbf{A}_{m,m^{\prime}}\left( w\left[ m^{\prime} \right] +\frac{\beta}{\alpha}w^*\left[ m^{\prime} \right] \right)},
\end{equation}
where $\mathbf{A}_{m,m^{\prime}}=e^{-j2\pi \left( c_2m^2+c_1{m^{\prime}}^2+m^{\prime}m/N \right)}/\sqrt{N}$. Considering that $\tilde{w}\left[ m \right] =w\left[ m \right] +\frac{\beta}{\alpha}w^*\left[ m \right] $ and  $w\left[ m \right] \sim \mathcal{C} \mathcal{N} \left( 0,\sigma _{n}^{2},0 \right) $, due to the presence of the conjugate term, $\tilde{w}\left[ m \right] $ will no longer satisfy the circularly symmetric property. Assume that $\tilde{w}\left[ m \right] \sim \mathcal{C} \mathcal{N} \left( \mu _{\tilde{w}},\sigma _{\tilde{w}}^{2},\kappa _{\tilde{w}} \right) $, then the mean of $\tilde{w}\left[ m \right] $ is
\begin{equation}
    \mu _{\tilde{w}}=\mathbb{E} \left\{ w\left[ m \right] \right\} +\frac{\beta}{\alpha}\mathbb{E} \left\{ w^*\left[ m \right] \right\} =0,
\end{equation}
where $\mathbb{E} \left\{ \cdot \right\} $ denotes the mathematical expectation operation, the variance of $\tilde{w}\left[ m \right] $ is
\begin{equation}
    \begin{aligned}
        \sigma _{\tilde{w}}^{2}&=\mathbb{E} \left\{ \left( w\left[ m \right] +\frac{\beta}{\alpha}w^*\left[ m \right] \right) \left( w\left[ m \right] +\frac{\beta}{\alpha}w^*\left[ m \right] \right) ^* \right\} 
\\
&=\left( 1+\left| \frac{\beta}{\alpha} \right|^2 \right) \mathbb{E} \left\{ \left| w\left[ m \right] \right|^2 \right\} +\left( \frac{\beta}{\alpha} \right) ^*\mathbb{E} \left\{ \left( w\left[ m \right] \right) ^2 \right\} 
\\ & \quad+\frac{\beta}{\alpha}\mathbb{E} \left\{ \left( w^*\left[ m \right] \right) ^2 \right\} 
\\
&=\left( 1+\left| \frac{\beta}{\alpha} \right|^2 \right) \sigma _{n}^{2},
    \end{aligned}
\end{equation}
and the pseudo-variance of $\tilde{w}\left[ m \right] $ is
\begin{equation}
    \begin{aligned}
        \kappa _{\tilde{w}}&=\mathbb{E} \left\{ \left( w\left[ m \right] \right) ^2+\left( \frac{\beta}{\alpha}w^*\left[ m \right] \right) ^2+2\frac{\beta}{\alpha}\left| w\left[ m \right] \right|^2 \right\} 
\\
&=\mathbb{E} \left\{ \left( w\left[ m \right] \right) ^2 \right\} +\left( \frac{\beta}{\alpha} \right) ^2\mathbb{E} \left\{ \left( w^*\left[ m \right] \right) ^2 \right\} \\&\quad+2\frac{\beta}{\alpha}\mathbb{E} \left\{ \left| w\left[ m \right] \right|^2 \right\} 
\\
&=2\frac{\beta}{\alpha}\sigma _{n}^{2}.
    \end{aligned}
\end{equation}

Since the interference term $\hat{v}\left[ m \right] $ is a linear combination of the random variables $\tilde{w}\left[ m \right] $, it also follows a complex Gaussian distribution, the numerical characteristics of $\hat{v}\left[ m \right] $ can be obtained according to the properties of the complex Gaussian distribution. Assume that $\hat{v}\left[ m \right] \sim \mathcal{C} \mathcal{N} \left( \mu _{\hat{v}},\sigma _{\hat{v}}^{2},\kappa _{\hat{v}} \right) $, then the mean of $\hat{v}\left[ m \right] $ is 
\begin{equation}
    \begin{aligned}
        \mu _{\hat{v}}=\sum_{m^{\prime}=0}^{N-1}{\mathbf{A}_{m,m^{\prime}}\mathbb{E} \left\{ \tilde{w}\left[ m^{\prime} \right] \right\}}=0,
    \end{aligned}
\end{equation}
the variance of $\hat{v}\left[ m \right] $ is
\begin{equation}
    \begin{aligned}
    \sigma _{\hat{v}}^{2} &= \sum_{m^{\prime}=0}^{N-1}\bigl| \mathbf{A}_{m,m^{\prime}} \bigr|^2\mathbb{E} \bigl\{ \bigl| \tilde{w}\bigl[ m^{\prime} \bigr] \bigr|^2 \bigr\} \\
    &= \sum_{m^{\prime}=0}^{N-1}\Bigl( \bigl| \frac{1}{\sqrt{N}}e^{-j2\pi \left( c_2m^2+c_1{m^{\prime}}^2+m^{\prime}m/N \right)} \bigr|^2 \\
    &\qquad \cdot \left( 1+\bigl| \frac{\beta}{\alpha} \bigr|^2 \right) \sigma _{n}^{2} \Bigr) \\
    &= \left( 1+\bigl| \frac{\beta}{\alpha} \bigr|^2 \right) \sigma _{n}^{2},
    \end{aligned}
\end{equation}
and the pseudo-variance of $\hat{v}\left[ m \right] $ is
\begin{equation}
    \begin{aligned}
        \kappa _{\hat{v}}&=\sum_{m^{\prime}=0}^{N-1}{\left( \mathbf{A}_{m,m^{\prime}} \right) ^2\mathbb{E} \left\{ \left( \tilde{w}\left[ m^{\prime} \right] \right) ^2 \right\}}
\\
&=\frac{1}{N}\sum_{m^{\prime}=0}^{N-1}{\left[ \left( e^{-j2\pi \left( c_2m^2+c_1{m^{\prime}}^2+\frac{m^{\prime}m}{N} \right)} \right) ^2\cdot 2\frac{\beta}{\alpha}\sigma _{n}^{2} \right]}
\\
&=2\sigma _{n}^{2}\cdot \frac{\beta}{\alpha}\cdot e^{-j2\pi \left( 2c_2m^2 \right)}\cdot \frac{1}{N}\sum_{m^{\prime}=0}^{N-1}{e^{-j2\pi \left( 2c_1{m^{\prime}}^2+2\frac{m^{\prime}m}{N} \right)}}
\\
&=2\sigma _{n}^{2}\cdot \frac{\beta}{\alpha}\cdot e^{-j2\pi \left( 2c_2m^2 \right)}\cdot \frac{1}{N}S\left( 2m \right) 
\\
&=\frac{2}{\sqrt{N}}\sigma _{n}^{2}\cdot \frac{\beta}{\alpha}\cdot e^{-j2\pi \left( 2c_2m^2 \right)}\cdot \left( 1-j \right) \cdot \gamma _{d,N}\cdot e^{j\frac{2\pi}{N}m^2d_{-1}}.
    \end{aligned}
\end{equation}
Therefore, $\hat{v}_{\mathrm{I}}\left[ m \right] $ and $\hat{v}_{\mathrm{Q}}\left[ m \right] $ respectively follow the Gaussian distributions $\hat{v}_{\mathrm{I}}\left[ m \right] \sim \mathcal{N} \left( 0,\sigma _{\hat{v},I}^{2} \right) $ and $\hat{v}_{\mathrm{Q}}\left[ m \right] \sim \mathcal{N} \left( 0,\sigma _{\hat{v},\mathrm{Q}}^{2} \right) $, where
\begin{subequations}
  \begin{align}
\sigma _{\hat{v},\mathrm{I}}^{2}&=\frac{1}{2}\left( \sigma _{\hat{v}}^{2}+\mathrm{Re}\left\{ \kappa _{\hat{v}} \right\} \right) ,\\
    \sigma _{\hat{v},\mathrm{Q}}^{2}&=\frac{1}{2}\left( \sigma _{\hat{v}}^{2}-\mathrm{Re}\left\{ \kappa _{\hat{v}} \right\} \right) .
  \end{align}
\end{subequations}
The above equation indicates that non-circular symmetry leads to differences in the equivalent noise power between the I and Q channels.

\subsection{Distribution of the Interference Component}
Given that the constellation points of $x\left[ m \right] $ are an energy-normalized M-QAM constellation with the same probability for each symbol, satisfying $\mathbb{E} \left\{ x\left[ m \right] \right\} =0$, $\mathbb{E} \{ \left| x\left[ m \right] \right|^2 \} =1$ and $\mathbb{E} \{ \left( x\left[ m \right] \right) ^2 \} =0$. Data symbols on different subcarriers are statistically independent.
The interference term $I_{\mathrm{ICI}}$ is a linear combination of the constellation points M-QAM. According to (\ref{eq:XI_mn}) and (\ref{eq:S_a}), the magnitude of the coefficient $\xi _m\left[ m^{\prime} \right] $ can only take two values of 0 and $\sqrt{2/N}$. 

When $N$ is sufficiently large, according to the central limit theorem, $I_{\mathrm{ICI}}$ can be approximated as following a complex Gaussian distribution. 
Assume that $I_{\mathrm{ICI}}\sim \mathcal{C} \mathcal{N} \left( \mu _{\mathrm{ICI}},\sigma _{\mathrm{ICI}}^{2},\kappa _{\mathrm{ICI}} \right) $, then the mean of $I_{\mathrm{ICI}}$ is
\begin{equation}
    \mu _{\mathrm{ICI}}=\frac{\beta}{\alpha}\sum_{
	m^{\prime}=0, 	m^{\prime}\ne m
}^{N-1}{\xi _m\left[ m^{\prime} \right] \mathbb{E} \left\{ x^*\left[ m^{\prime} \right] \right\} =0,}
\end{equation}
the variance of $I_{\mathrm{ICI}}$ is
\begin{equation}
    \begin{aligned}
        \sigma _{\mathrm{ICI}}^{2}&=\left| \frac{\beta}{\alpha} \right|^2\sum_{m^{\prime}=0, 	m^{\prime}\ne m}^{N-1}{\left| \xi _m\left[ m^{\prime} \right] \right|^2\mathbb{E} \left\{ \left| x^*\left[ m^{\prime} \right] \right|^2 \right\}}
\\
&=\left| \frac{\beta}{\alpha} \right|^2\sum_{m^{\prime}=0, 	m^{\prime}\ne m}^{N-1}{\left| \xi _m\left[ m^{\prime} \right] \right|^2}
\\ &=\left| \frac{\beta}{\alpha} \right|^2\frac{2}{N}\cdot \left( \frac{N}{2}-1 \right) ,
    \end{aligned}
\end{equation}
and the pseudo-variance of $I_{\mathrm{ICI}}$ is
\begin{equation}
    \kappa _{\mathrm{ICI}}=\left( \frac{\beta}{\alpha} \right) ^2\sum_{m^{\prime}=0, m^{\prime}\ne m}^{N-1}{\left( \xi _m\left[ m^{\prime} \right] \right) ^2\mathbb{E} \left\{ \left( x^*\left[ m^{\prime} \right] \right) ^2 \right\}}=0.
\end{equation}
Therefore, $I_{\mathrm{I}}^{\mathrm{ICI}},I_{\mathrm{Q}}^{\mathrm{ICI}}\sim \mathcal{N} ( 0,\frac{N-2}{2N}\left| \frac{\beta}{\alpha} \right|^2 ) $.

\section{Error Rate Analysis}
\begin{figure}
    \centering
    \includegraphics[width=0.75\linewidth]{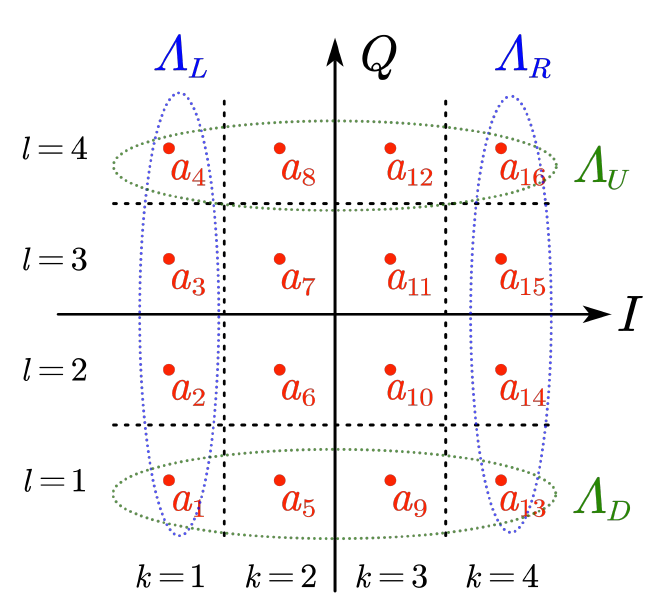}
    \caption{Diagram of 16QAM constellation point labeling.}
    \label{fig:16QAM_cons}
\end{figure}

In this section, we analyze the BER performance of AFDM systems under IQ imbalance at the receiver and derive a closed-form expression. 
Let the symbol error rate of the M-QAM constellation be expressed as follows
\begin{equation}
    P_S=1-P_C,
\end{equation}
where $P_C\triangleq \left( 1-P_{\mathrm{I}} \right) \left( 1-P_{\mathrm{Q}} \right) $ is the probability of the correct symbol decision of the M-QAM system, while $P_{\mathrm{I}}$ and $P_{\mathrm{Q}}$ are the probability of the error symbol decision of the $\sqrt{M}$-level PAM signals for the I-channel and the Q-channel, respectively.
When all symbols are equally probable and Gray coding is used, the bit error rate is
\begin{equation}
    P_b\approx \frac{1}{\log _2M}P_S=\frac{1}{\log _2M}\left( 1-P_C \right) .
    \label{eq:BER}
\end{equation}
In the following, $P_C$ is calculated for different symbol values. The energy-normalized M-QAM constellation points can be represented as follows
\begin{equation}
    a_i=\frac{1}{\sqrt{\frac{2}{3}\left( M-1 \right)}}\left[ \left( 2k-1-\sqrt{M} \right) +j\left( 2l-1-\sqrt{M} \right) \right],
\end{equation}
where $k,l\in \left\{ 1,2,\cdots ,\sqrt{M} \right\} $, $M$ denotes the modulation order and the constellation point index $i=l+\sqrt{M}\left( k-1 \right) $. Taking 16QAM as an example, the constellation point labels are shown in Fig. \ref{fig:16QAM_cons}.
The minimum distance between constellation points is $D=\sqrt{6/\left( M-1 \right)}$.

Consider the I channel. Given the different constellation points, three cases are considered: $k=1$, $k=\sqrt{M}$, and $k=2,\cdots ,\sqrt{M}-1$. For the constellation point set $\varLambda _L$ with $k=1$, the error probability of the I-channel symbol when transmitting symbol $a_i$ can be expressed as
\begin{equation}
    \begin{aligned}
        &P_{\mathrm{I}}\left( e\mid x\left[ m \right] =a_i,a_i\in \varLambda _L \right) 
\\
\,\, &\quad=\mathrm{Pr}\left\{ \hat{y}_{\mathrm{I}}\left[ m \right] >a_{\mathrm{I}}+\frac{D}{2} \right\} 
\\
&\quad=\mathrm{Pr}\left\{ e_{\mathrm{I}}+I_{\mathrm{I}}^{\mathrm{ICI}}+\hat{v}_{\mathrm{I}}\left[ m \right] >\frac{D}{2} \right\} 
\\
\,\, &\quad=\mathrm{Pr}\left\{ \frac{I_{\mathrm{I}}^{\mathrm{ICI}}+\hat{v}_{\mathrm{I}}\left[ m \right]}{\sqrt{\frac{1}{2}\sigma _{\mathrm{ICI}}^{2}+\sigma _{\hat{v},\mathrm{I}}^{2}}}>\frac{\frac{D}{2}-e_{\mathrm{I}}}{\sqrt{\frac{1}{2}\sigma _{\mathrm{ICI}}^{2}+\sigma _{\hat{v},\mathrm{I}}^{2}}} \right\} 
\\
&\quad=Q\left( \frac{\frac{D}{2}-e_{\mathrm{I}}}{\sqrt{\frac{1}{2}\sigma _{\mathrm{ICI}}^{2}+\sigma _{\hat{v},\mathrm{I}}^{2}}} \right) .
    \end{aligned}
\end{equation}
Similarly, for the constellation point set $\varLambda _R$ with $k=\sqrt{M}$,
\begin{equation}
    P_{\mathrm{I}}\left( e\mid x\left[ m \right] =a_i,a_i\in \varLambda _R \right) =Q\left( \frac{\frac{D}{2}+e_{\mathrm{I}}}{\sqrt{\frac{1}{2}\sigma _{\mathrm{ICI}}^{2}+\sigma _{\hat{v},\mathrm{I}}^{2}}} \right),
\end{equation}
and for the other constellation points where $k=2,\cdots ,\sqrt{M}-1$,
\begin{equation}
    \begin{aligned}
        &P_{\mathrm{I}}\left( e\mid x\left[ m \right] =a_i,a_i\notin \varLambda _L,a_i\notin \varLambda _R \right) 
\\
&\quad=Q\left( \frac{\frac{D}{2}+e_{\mathrm{I}}}{\sqrt{\frac{1}{2}\sigma _{\mathrm{ICI}}^{2}+\sigma _{\hat{v},\mathrm{I}}^{2}}} \right) +Q\left( \frac{\frac{D}{2}-e_{\mathrm{I}}}{\sqrt{\frac{1}{2}\sigma _{\mathrm{ICI}}^{2}+\sigma _{\hat{v},\mathrm{I}}^{2}}} \right) .
    \end{aligned}
\end{equation}
In the aforementioned expression for the symbol error probability of the I channel, $e_{\mathrm{I}}$ is related to the subcarrier index and the constellation point value, $\sigma _{\hat{v},\mathrm{I}}^{2}$ is related only to the subcarrier index, and $\sigma _{\mathrm{ICI}}^{2}$ is independent of both the constellation point value and the subcarrier index.
Denote the symbol error probability of the $m$-th subcarrier when transmitting the symbol $a_i$ as $P_{\mathrm{I}}^{\left( i \right)}\left[ m \right] $, the value of $e_{\mathrm{I}}$ for the $m$-th subcarrier when transmitting the symbol $a_i$ as $e_{\mathrm{I}}^{\left( i \right)}\left[ m \right] $, and the value of $\sigma _{\hat{v},\mathrm{I}}^{2}$ for the $m$-th subcarrier as $\sigma _{\hat{v},\mathrm{I}}^{2}\left[ m \right] $. Then
\begin{equation}
\begin{aligned}
        &P_{\mathrm{I}}^{\left( i \right)}\left[ m \right] \\
        &=\begin{cases}
	Q\left( \frac{\frac{D}{2}-e_{\mathrm{I}}^{\left( i \right)}\left[ m \right]}{\sqrt{\frac{1}{2}\sigma _{\mathrm{ICI}}^{2}+\sigma _{\hat{v},\mathrm{I}}^{2}\left[ m \right]}} \right) ,	a_i\in \varLambda _L\\
	Q\left( \frac{\frac{D}{2}+e_{\mathrm{I}}^{\left( i \right)}\left[ m \right]}{\sqrt{\frac{1}{2}\sigma _{\mathrm{ICI}}^{2}+\sigma _{\hat{v},\mathrm{I}}^{2}\left[ m \right]}} \right) ,	a_i\in \varLambda _R\\
	Q\left( \frac{\frac{D}{2}-e_{\mathrm{I}}^{\left( i \right)}\left[ m \right]}{\sqrt{\frac{1}{2}\sigma _{\mathrm{ICI}}^{2}+\sigma _{\hat{v},\mathrm{I}}^{2}\left[ m \right]}} \right)+Q\left( \frac{\frac{D}{2}+e_{\mathrm{I}}^{\left( i \right)}\left[ m \right]}{\sqrt{\frac{1}{2}\sigma _{\mathrm{ICI}}^{2}+\sigma _{\hat{v},\mathrm{I}}^{2}\left[ m \right]}} \right),	else.\\
\end{cases}
\end{aligned}
\end{equation}
Similarly, for the Q channel, the symbol error probability $P_{\mathrm{Q}}^{\left( i \right)}\left[ m \right] $ for the $m$-th subcarrier when transmitting the symbol $a_i$ is given by
\begin{equation}
    \begin{aligned}
        &P_{\mathrm{Q}}^{\left( i \right)}\left[ m \right] \\
        &=\begin{cases}
	Q\left( \frac{\frac{D}{2}-e_{\mathrm{Q}}^{\left( i \right)}\left[ m \right]}{\sqrt{\frac{1}{2}\sigma _{\mathrm{ICI}}^{2}+\sigma _{\hat{v},\mathrm{Q}}^{2}\left[ m \right]}} \right) ,a_i\in \varLambda _D\\
	Q\left( \frac{\frac{D}{2}+e_{\mathrm{Q}}^{\left( i \right)}\left[ m \right]}{\sqrt{\frac{1}{2}\sigma _{\mathrm{ICI}}^{2}+\sigma _{\hat{v},\mathrm{Q}}^{2}\left[ m \right]}} \right) ,a_i\in \varLambda _U\\
	Q\left( \frac{\frac{D}{2}-e_{\mathrm{Q}}^{\left( i \right)}\left[ m \right]}{\sqrt{\frac{1}{2}\sigma _{\mathrm{ICI}}^{2}+\sigma _{\hat{v},\mathrm{Q}}^{2}\left[ m \right]}} \right) +Q\left( \frac{\frac{D}{2}+e_{\mathrm{Q}}^{\left( i \right)}\left[ m \right]}{\sqrt{\frac{1}{2}\sigma _{\mathrm{ICI}}^{2}+\sigma _{\hat{v},\mathrm{Q}}^{2}\left[ m \right]}} \right) ,else.\\
\end{cases}
    \end{aligned}
\end{equation}

Therefore, the symbol correct decision probability of the M-QAM system for the $m$-th subcarrier when transmitting symbol $a_i$ is
\begin{equation}
    P_{C}^{\left( i \right)}\left[ m \right] =\left( 1-P_{\mathrm{I}}^{\left( i \right)}\left[ m \right] \right) \left( 1-P_{Q}^{\left( i \right)}\left[ m \right] \right) .
\end{equation}
After averaging over different subcarriers and symbol values, the symbol correct decision probability is obtained as
\begin{equation}
    P_C=\frac{1}{MN}\sum_{m=0}^{N-1}{\sum_{i=1}^M{\left( 1-P_{\mathrm{I}}^{\left( i \right)}\left[ m \right] \right) \left( 1-P_{Q}^{\left( i \right)}\left[ m \right] \right)}}.
\end{equation}
Finally, substituting $P_C$ into equation (\ref{eq:BER}) produces the system BER.

\section{Numerical Results}
\begin{figure}
    \centering
    \includegraphics[width=1\linewidth]{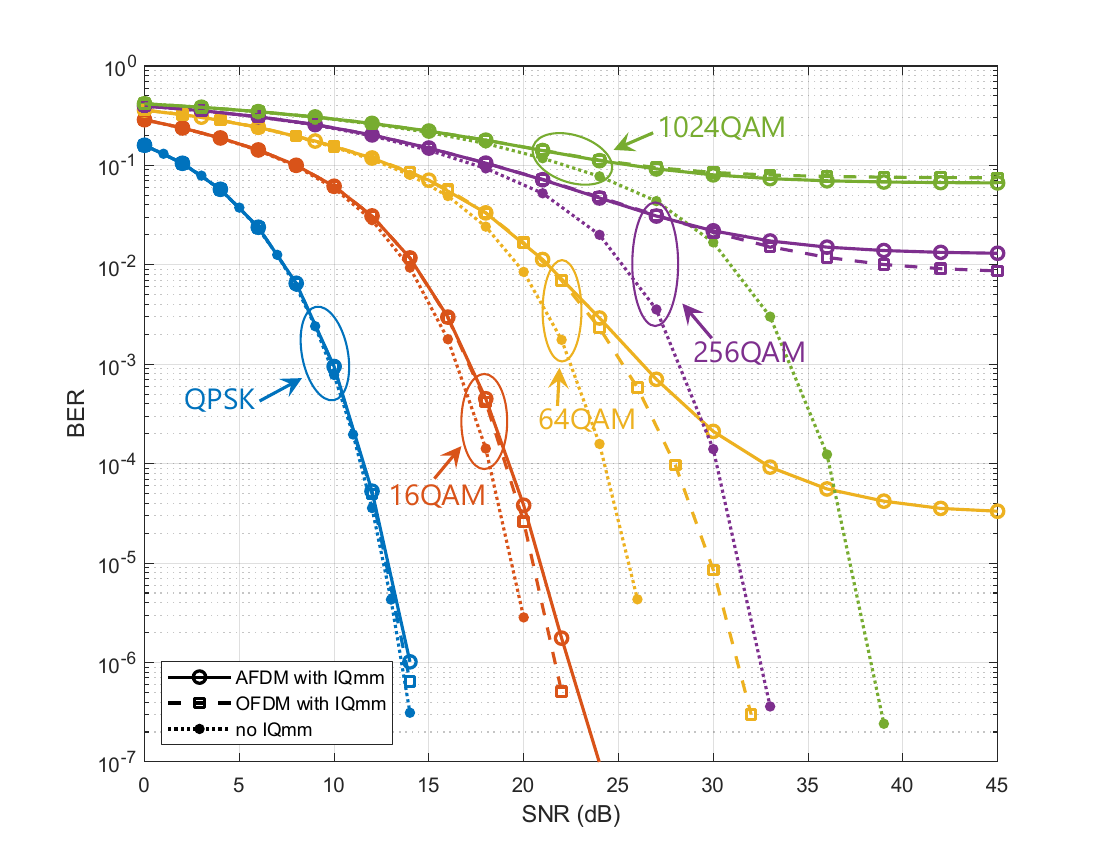}
    \caption{The BER performance of AFDM and OFDM under IQ imbalance at the receiving end ($\varepsilon =5\%,\varphi =1.5\degree$).}
    \label{fig:AFDM_OFDM_BERcomp_typ}
\end{figure}

In this section, we perform numerical simulations of the BER performance of AFDM systems under IQ effects at the receiver using the Monte-Carlo method. 
Considering that several typical IQ demodulators provide IQ amplitude imbalance and phase imbalance parameters in their data sheets \cite{ADL5375}, \cite{ADMV4540}, \cite{LTC5594}, as shown in Table \ref{tab:typ_val}, we select the amplitude imbalance parameter $\varepsilon =5\%$ and the phase imbalance parameter $\varphi =1.5\degree $ as representative values to evaluate the BER performance of AFDM under the influence of receiver-side IQ imbalance. 
This performance is then compared with the BER performance of OFDM systems under the same conditions, with the results shown in Fig. \ref{fig:AFDM_OFDM_BERcomp_typ}.
The number of subcarriers used in the test is $N=256$.
\begin{table}
    \centering
\caption{Performance parameters of IQ imbalance for several typical IQ demodulators.}
\label{tab:typ_val}
    \begin{tabular}{cccc}\toprule
         Model of IQ Demodulator&  $\epsilon$ (dB)&  $\epsilon$ (\%)& $\varphi$ (deg)\\\midrule
         ADL5375-05&  0.07dB
&  0.81\%& 1.7°
\\
         ADL5375-15
&  0.10dB
&  1.16\%& 1.49°
\\
         ADMV4540
&  0.5dB
&  5.93\%& 1.6°
\\
         LTC5594
&  0.44dB
&  5.20\%& 1.0°
\\ \bottomrule
    \end{tabular}

\end{table}

Simulation results show that under the same IQ imbalance parameter, higher-order modulation is more severely affected by IQ imbalance than lower-order modulation. 
This occurs because the constellation point spacing $D$ is smaller for higher modulation orders, making them more vulnerable to IQ imbalance for both OFDM and AFDM. 
This phenomenon can also be observed in the expression for the theoretical BER value.

\begin{figure}[!t]
    \centering
    % 子图(a)：上图
    \begin{subfigure}[b]{\linewidth}  % 子图宽度占满当前行宽（100%）
        \centering
        \includegraphics[width=0.9\linewidth]{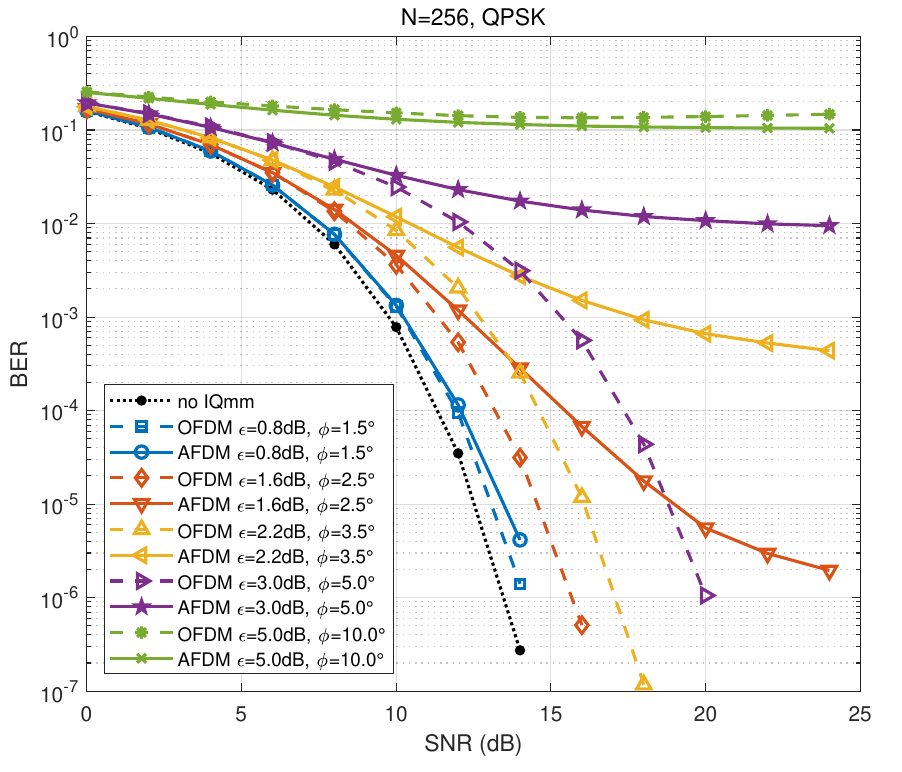}  % 图片宽度可自定义（避免过高）
        \caption{QPSK}
        \label{subfig:QPSK}
    \end{subfigure}
    \\[6pt]  % 换行并添加6pt间距（上下排列必备，避免重叠）
    % 子图(b)：下图
    \begin{subfigure}[b]{\linewidth}
        \centering
        \includegraphics[width=0.9\linewidth]{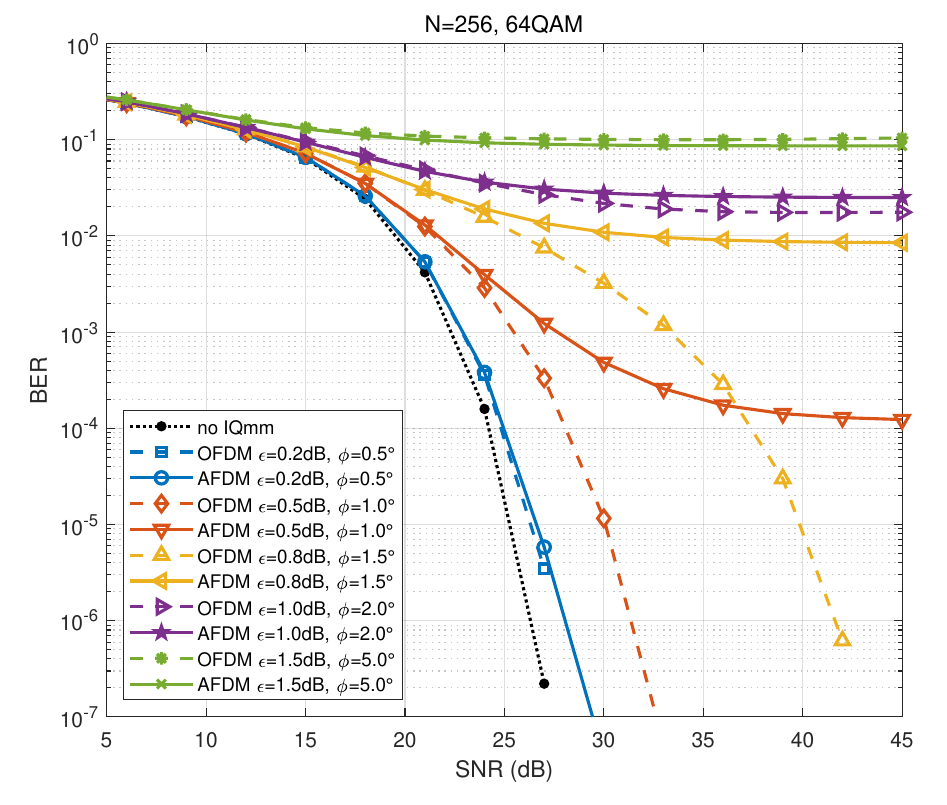}
        \caption{64QAM}
        \label{subfig:64QAM}
    \end{subfigure}
    % 总图标题
    \caption{Comparison of BER performance between AFDM and OFDM under different IQ imbalance cases.}
    \label{fig:AFDM_OFDM_BERcomp_QAM}
\end{figure}

In Fig. \ref{fig:AFDM_OFDM_BERcomp_QAM}, the differences in BER between AFDM and OFDM at various QAM modulation orders are tested as a function of IQ imbalance severity. The simulation results indicate that, under most conditions, AFDM exhibits worse BER performance than OFDM for identical IQ imbalance parameters. Only under extremely severe IQ imbalance conditions, which rarely occur in practical engineering scenarios, might OFDM demonstrate worse performance than AFDM.
For minor IQ imbalance, the BER decreases sharply with increasing SNR, following the asymptotic behavior of the Q-function. However, in the presence of severe IQ imbalance, the BER curve departs from this trend at high SNR and exhibits a saturation effect, converging to a stable non-zero value, resulting in an error floor.

This phenomenon arises from the ICI induced by the receiver IQ imbalance. Specifically, compared to the ideal case without IQ imbalance, the received constellation points are displaced. Once this offset crosses the symbol decision thresholds, symbol errors become inevitable. Consequently, with increasing SNR, the error probability for these symbols converges to 1. 
Under this condition, the asymptotic BER is governed by the proportion of symbol realizations for which the displaced constellation points exceed the decision thresholds among all possible symbol values. 

Typically, for OFDM, interference terms originate solely from another subcarrier, with the offset having at most $M$ possible values and a relatively small maximum offset. However, for AFDM, interference terms represent the weighted sum of data symbols across other subcarriers, featuring a larger maximum offset and a higher probability of exceeding the decision threshold compared to OFDM. Consequently, AFDM exhibits poorer BER performance than OFDM.

\begin{figure}
    \centering
    \includegraphics[width=1\linewidth]{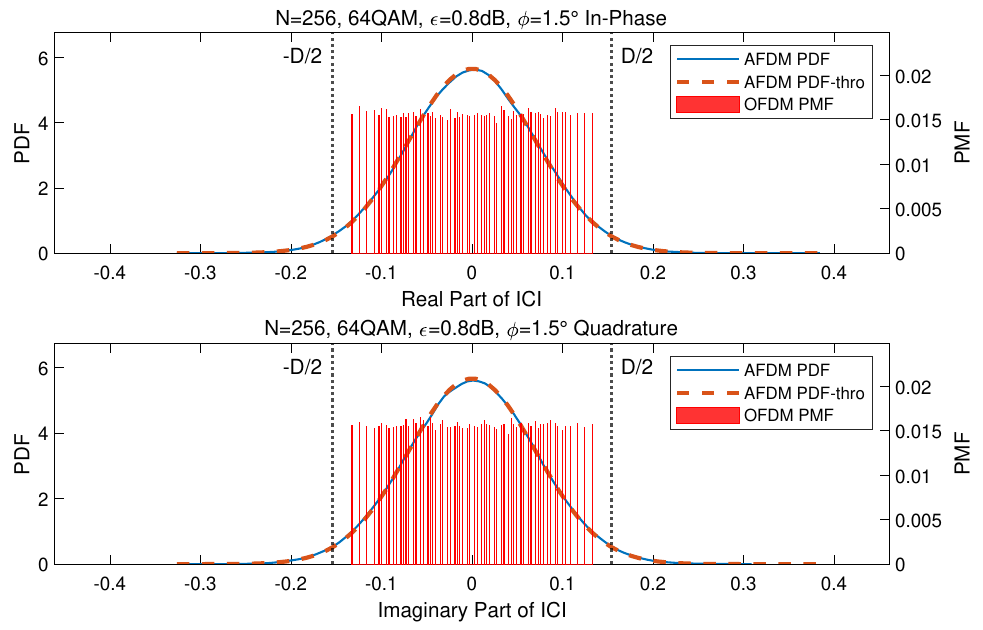}
    \caption{The probability distribution of the interference component $I_{\mathrm{ICI}}$ in I and Q channels.}
    \label{fig:ICI_compare}
\end{figure}

For example, for the case with an IQ imbalance parameter of $\varepsilon =0.8\mathrm{dB},\varphi =1.5\degree $ in Fig. \ref{subfig:64QAM}, the probability distributions of the I and Q components of the ICI term on a single subcarrier in (\ref{eq:AFDM_IQmm_y_eleform_sep}) are separately simulated. The results are shown in Fig. \ref{fig:ICI_compare}. 
The solid blue line represents the probability density function (PDF) of the interference component $I_{\mathrm{ICI}}$ in AFDM, obtained by numerical simulation statistics. The dashed orange line depicts the Gaussian distribution PDF, plotted under the assumption of circular symmetry and complex Gaussian approximation $I_{\mathrm{ICI}}\sim \mathcal{C} \mathcal{N} ( 0,\frac{N-2}{N}\left| \frac{\beta}{\alpha} \right|^2,0 ) $. The red bar chart shows the discrete probability mass function (PMF) of the interference component $I_{\mathrm{ICI}}$ in OFDM, derived from numerical simulation statistics. The black vertical dashed line indicates the current decision threshold $D/2$.
The solid blue line perfectly aligns with the dashed orange curve, indicating that approximating the interference term in AFDM as a circularly symmetric complex Gaussian distribution is reasonable.
Under the current IQ imbalance parameters, all possible values of the OFDM interference term remain below the decision threshold, whereas a significant portion of the possible values for the AFDM interference term may exceed the decision threshold. Consequently, as the SNR increases, the BER of OFDM rapidly decreases below $10^{-6}$, while the BER of AFDM remains at the $10^{-2}$ level. The difference in the interference term distribution between AFDM and OFDM under IQ imbalance at the receiver end accounts for their performance disparity.

\begin{figure}[!t]
    \centering
    % 子图(a)：上图
    \begin{subfigure}[b]{\linewidth}  % 子图宽度占满当前行宽（100%）
        \centering
        \includegraphics[width=0.9\linewidth]{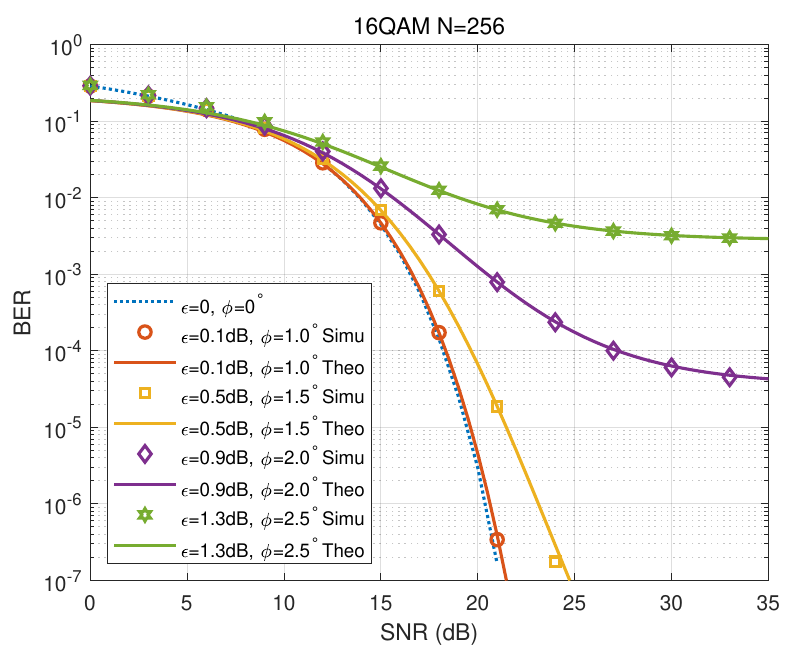}  % 图片宽度可自定义（避免过高）
        \caption{16QAM}
        \label{subfig:16QAM}
    \end{subfigure}
    \\[6pt]  % 换行并添加6pt间距（上下排列必备，避免重叠）
    % 子图(b)：下图
    \begin{subfigure}[b]{\linewidth}
        \centering
        \includegraphics[width=0.9\linewidth]{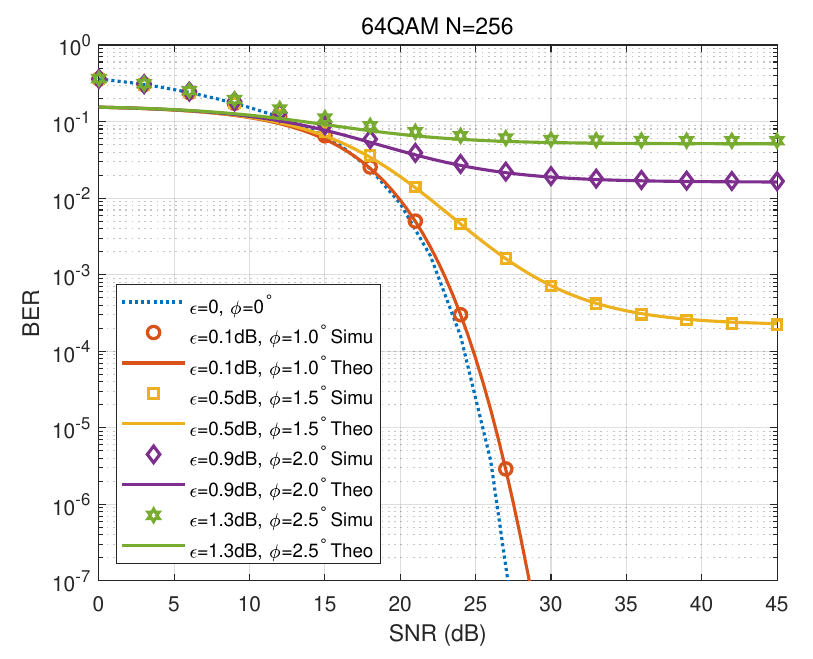}
        \caption{64QAM}
        \label{subfig:64QAM}
    \end{subfigure}
    % 总图标题
    \caption{Comparison of theoretical and simulated BER performance of AFDM under IQ imbalance at receiver.}
    \label{fig:AFDM_BERthro}
\end{figure}

The fixed number of carrier carriers and modulation order were maintained while observing the impact of different IQ imbalance parameters on BER. The simulation results were compared with the theoretical BER values, as shown in Fig. \ref{fig:AFDM_BERthro}. The points in the figure represent the simulated BER results, while the solid line indicates the theoretical BER values. The results demonstrate excellent agreement between the theoretical and simulated values, validating the proposed BER theory.

\section{Conclusion}
This paper has conducted a systematic modeling and theoretical analysis of the impact of receiver IQ imbalance on the performance of AFDM systems. A mathematical model of the AFDM transceiver under receiver IQ imbalance is established, revealing the intrinsic mechanism by which IQ imbalance induces ICI in AFDM systems. 
Furthermore,  a closed-form BER expression is derived for M-QAM AFDM systems under receiver IQ imbalance over AWGN channels and validated through numerical simulations.
% Numerical results demonstrate that, under identical IQ imbalance conditions, the BER performance of AFDM systems is generally inferior to that of OFDM systems. 
% This behavior fundamentally stems from the fact that, unlike OFDM where interference originates only from a single image subcarrier, the IQ imbalance in AFDM introduces aggregated interference from multiple subcarriers, which significantly increases the probability that the received constellation points cross the decision thresholds, thereby causing the BER error floor to appear at relatively lower signal-to-noise ratios. 
The results of this study indicate that, although AFDM exhibits several advantages in high-mobility channel environments, it is more sensitive to receiver IQ imbalance compared to OFDM. In particular, for high-order modulation schemes, even a relatively small degree of IQ imbalance in practical receiver hardware can lead to non-negligible performance degradation in AFDM systems. 
Future work will focus on investigating the impact of IQ imbalance on the diversity performance of AFDM over doubly selective channels, developing joint estimation and compensation algorithms for IQ imbalance, and extending the analysis to account for other hardware impairments, such as phase noise, in AFDM systems.

\ifCLASSOPTIONcaptionsoff
  \newpage
\fi



% \bibliographystyle{IEEEtran}
% \bibliography{IEEEabrv,../bib/paper}

\bibliographystyle{IEEEtran} % IEEE标准参考文献样式
% 路径改为根目录的paper.bib（无需写后缀，IEEEabrv保留）
\bibliography{bibtex/bib/references}



% \end{thebibliography}


% \end{IEEEbiography}


%\newpage



%\vfill


\end{document}